\documentclass[fleqn,usenatbib]{mnras}
\usepackage{newtxtext,newtxmath}
\usepackage[T1]{fontenc}
\DeclareRobustCommand{\VAN}[3]{#2}
\let\VANthebibliography\thebibliography
\def\thebibliography{\DeclareRobustCommand{\VAN}[3]{##3}\VANthebibliography}

\usepackage{graphicx}
\usepackage{amsmath}
\usepackage{bm}
\usepackage{natbib}
\usepackage{siunitx}
\usepackage{gensymb}
\usepackage{url}
\usepackage{float}
\usepackage{multicol}
\usepackage{pdflscape}

\title[Umbral Oscillations]{Chromospheric resonator model for sunspot revealed by multi-height observation of umbral wave}

\author[Sangal et al.]{
Kartika Sangal,$^{1,2}$
A.K.~Srivastava,$^{3}$
Libo Fu,$^{1,2}$
Ding~Yuan,$^{1,2}$\thanks{E-mail: yuanding@hit.edu.cn}
Song Feng,$^{4}$
and Yuandeng Shen$^{1,2}$
\\
$^{1}$Key Laboratory of Solar Activity and Space Weather, School of Aerospace, Harbin Institute of Technology, Shenzhen, Guangdong, China\\
$^{2}$Shenzhen Key Laboratory of Numerical Prediction for Space Storm, School of Aerospace, Harbin Institute of Technology, Shenzhen, Guangdong, China\\
$^{3}$Department of Physics, Indian Institute of Technology (BHU), Varanasi-221005, India\\
$^{4}$ Faculty of Information Engineering and Automation, Kunming University of Science and Technology,Kunming 650500,  China
}

\date{Accepted 2026 May 13. Received 2026 May 13; in original form 2026 January 17}

\pubyear{\the\year{}}

\begin{document}
\label{firstpage}
\pagerange{\pageref{firstpage}--\pageref{lastpage}}
\maketitle

\begin{abstract}
Sunspots are transient, magnetically intense features that host oscillations linked to magnetohydrodynamic (MHD) waves. These waves may contribute to plasma heating and drive mass flows in the solar wind. Beyond their energetic role, they serve as diagnostic tools for probing sunspot structure. In this study, we investigated chromospheric wave propagation in a sunspot using high-resolution, multi-wavelength observations from the Goode Solar Telescope at Big Bear Solar Observatory. Spectral analysis shows that the intensity at H$\alpha$ line core and its wings exhibited oscillatory signal at about 3 min. We performed a cross-wavelet analysis to examine the phase relationship between the wing-integrated and line-core intensity oscillations of the H$\alpha$ line and the centroid-derived H$\alpha$ Doppler velocity. We also analyze the phase relationships between intensity pairs from different passband combinations of the H$\alpha$ line. The results indicate the presence of slow magnetoacoustic modes manifesting standing waves along with upward propagating waves. The observed phase patterns suggest that umbral waves are confined within a non-ideal acoustic resonator, providing measurable wave properties that could serve as input for sunspot seismology and refine models of sunspot atmospheric structure.

\end{abstract}




\section{Introduction} \label{sec:intro}

Sunspots are regions of intense magnetic fields that appear at the solar photosphere. They typically consist of a dark central umbra surrounded by a penumbra. These magnetic structures host a variety of oscillations and waves. Sunspot oscillations have been studied extensively due to their important role in solar atmospheric dynamics \citep[e.g.,][]{1969SoPh....7..351B,2000SoPh..192..373B,2006RSPTA.364..313B,2015LRSP...12....6K}. Sunspot waves serve as a primary source of upward-propagating disturbances that extend into the chromosphere, transition region, and corona. As they propagate, these waves influence energy transport, drive mass flows, and contribute to plasma heating \citep[e.g.,][]{2010ApJ...724L.194V,2012ApJ...757..160J,2014A&A...561A..19Y,2017ApJ...836...18C,2017ApJ...847....5K,2018ApJ...855...65H}. During upward propagation, the waves can steepen into shocks, dissipating their energy and potentially providing a source of heating for the chromosphere or corona \citep[e.g.,][]{2006ApJ...640.1153C,2014ApJ...786..137T}. These oscillations not only reflect the generation of acoustic energy beneath the solar surface but are also shaped by the magnetic and thermal structure of the sunspot. Consequently, their spatial and spectral characteristics provide valuable diagnostic information, allowing us to probe the internal structure of sunspots and understand the surrounding plasma environment \citep[e.g.,][]{2011ApJ...728...84B,2016A&A...594A.101Y,2022RAA....22k5009F}.

Oscillations in sunspots are observed at multiple characteristic periods. In the photosphere, five-minute oscillations are dominant with an average rms velocity amplitude of few hundred $\mathrm{m\,s^{-1}}$ \citep{1982Natur.297..485T, 1986ApJ...311.1015A}. These oscillations are driven by global p-mode acoustic waves. In contrast, three-minute oscillations become increasingly prominent in the upper layers of the solar atmosphere and have substantially higher velocity amplitudes of a few kilometers per second \citep{1998ApJ...497..464L}. These oscillations are interpreted as slow magnetoacoustic waves that propagate along magnetic field lines \citep[e.g.,][]{2006ApJ...640.1153C,2013ApJ...779..168J,2015LRSP...12....6K}. Three-minute oscillations were first observed as "umbral flashes", appearing as brightness fluctuations in the Ca II K emission line above the sunspot umbra \citep[e.g.,][]{1969SoPh....7..351B}. They have since been detected as intensity variations in the Ca II H and K lines, as well as velocity shifts in spectral lines such as He I 10830 \AA. Although these oscillations can occur at multiple heights in the solar atmosphere, they are most clearly observed in the chromosphere, transition region, and corona above sunspot umbra. Numerous studies have confirmed that three-minute oscillations dominate the chromospheric umbra \citep[e.g.,][]{1969SoPh....7..351B,1982ApJ...253..386L,1987SoPh..108...61G,2006RSPTA.364..313B,2007PASJ...59S.631N}.

Two main theories have been proposed to explain the three-minute oscillations observed in sunspot umbra. The first is based on resonant cavities for magnetoacoustic waves within the vertically stratified umbral atmosphere \citep[e.g.,][]{1981SoPh...71...21S,1982SoPh...79...19T,1985ApJ...294..682L,1987SoPh..108...61G,2008SoPh..251..501Z}. These cavities are classified according to their location in the solar atmosphere. In the photospheric resonance model, the cavity lies within the photosphere and subphotosphere. In this case, oscillations can be driven by overstable convective regions in the umbral photosphere. They can also result from selective transmission of high-frequency components from surrounding p-mode oscillations. The chromospheric resonance model, on the other hand, confines the cavity between the temperature minimum and the steep temperature gradient in the upper chromosphere and transition region. Here, waves are trapped between a lower boundary defined by reflection at the high cut-off frequency near the temperature minimum and an upper boundary formed by the sudden increase in sound speed in the transition region \citep{1981NASSP.450..263L}. An alternative theory is based on wave propagation in a stratified solar atmosphere. In this model, the combined effects of stratification and the acoustic cut-off frequency lead to the observed oscillations without requiring a cavity \citep[e.g.,][]{2010ApJ...722..131F,2015ApJ...802...45C,2017ApJ...847....5K}. High-frequency waves, which lie above the cut-off, are able to propagate upward. Lower-frequency waves, on the other hand, are reflected back. This reflection acts as a natural filtering mechanism, allowing only certain waves to reach higher layers of the atmosphere.

Resonances in both the photosphere \citep{2019ApJ...883...72C} and the chromosphere of sunspot umbra have received considerable attention in recent years \citep[e.g.,][]{2008SoPh..251..501Z,2011ApJ...728...84B,2015A&A...580A.107S,2016ApJS..224...30Y,2018MNRAS.481..262W,2019A&A...627A.169F}. Numerical simulations, including those based on ideal MHD \citep{2011ApJ...728...84B,2015A&A...580A.107S} and two-fluid models \citep{2018MNRAS.481..262W}, demonstrate that resonant cavities can sustain and amplify three-minute oscillations. Observational evidence supports this interpretation as well. For instance, \citet{2020NatAs...4..220J} reported a distinct high-frequency power peak in He I 10830 \AA\ above sunspots, consistent with the presence of a chromospheric cavity. At the same time, the acoustic cut-off frequency continues to shape the vertical propagation and frequency distribution of waves, even in the presence of a resonator \citep{2019A&A...627A.169F}. Hence, both resonant cavity effects and atmospheric stratification contribute to the three-minute oscillations in sunspot umbra.

In this study, we analyze observational data from the BBSO to investigate the source of three-minute oscillations in a sunspot umbra. Section \ref{sec: data} describes the observations and data analysis. Section \ref{sec:methodology} outlines the methodology applied. The results are presented in Section \ref{sec:results}. Finally, Section \ref{sec:discussion} discusses these findings and draws the main conclusions.

\begin{figure*}
\centering
\includegraphics[width=.90\linewidth]{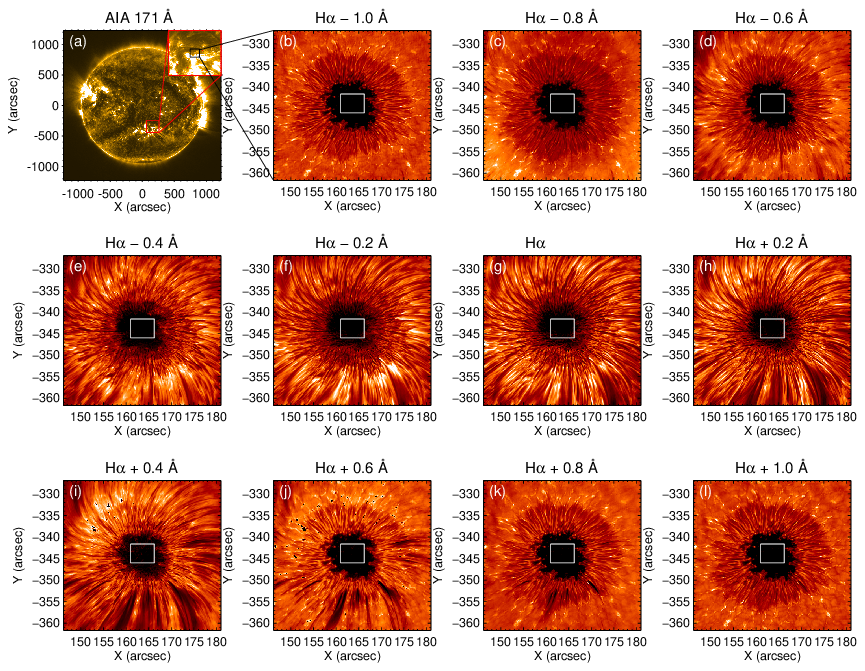}
\caption{Multi-instrument observations of the active region. (a) Full-disk solar image from AIA/SDO in the 171 \AA\ filter. The red box indicates the region of interest. On the top right of the panel, zoomed in view of the red box is shown,and within which a smaller black box is overplotted, marking the GST field of view. (b-f) Zoomed in view of the selected region in the H$\alpha$ blue wings. (g) Zoomed in view in the H$\alpha$ line core. (h-l) Zoomed in view in the H$\alpha$ red wings. In all panels, the over plotted white box indicates the subregion chosen for analysis.}
\label{fig:fig1}
\end{figure*}

\section{Observations and Data Analysis} 
\label{sec: data}
\subsection{BBSO data processing} 

We analyzed NOAA AR 12384 using observational data from the Visible Imaging Spectrometer (VIS) installed on the Goode Solar Telescope (GST) at the Big Bear Solar Observatory (BBSO) \citep[e.g.,][]{2010AN....331..636C}. The observations were carried out on July 14, 2015, between 17:36 and 18:06 UT, with a total duration of about 30 minutes. The GST provides a maximum field of view (FOV) of 70$^{\prime\prime}\times$70$^{\prime\prime}$ and achieves a spatial resolution of 0.2$^{\prime\prime}$\ at 1.56 \micro m and 0.06$^{\prime\prime}$\ at 0.5 \micro m \citep[e.g.,][]{2010AN....331..636C,2023NatAs...7..856Y}. The VIS operates in the wavelength range of 5500–7000 \AA\ and uses a Fabry-P\'erot etalon to provide a narrow bandpass of 0.07 \AA\ across a circular FOV of 70$^{\prime\prime}$ in diameter. It is capable of observing several key spectral lines, including H$\alpha$ 6563 \AA, Fe I 6300 \AA, and Na I D2 5890 \AA.

The sunspot was scanned in the H$\alpha$ line at 6563~\AA\ in steps of 0.2 \AA, covering both the blue wing ($-1.0$ \AA) and the red wing ($+1.0$ \AA). Chromospheric images were obtained with a cadence of 19 seconds. All frames were speckle reconstructed to correct for atmospheric turbulence. The observed region is outlined by a black box in the zoomed view shown in the top right corner of the AIA 171 \AA\ image (Fig.~\ref{fig:fig1}(a)). The other panels show zoomed in view of the H$\alpha$ line core and wings, revealing the umbra and the surrounding penumbra. To process the temporal image sequence, we averaged each 3×3-pixel region, reducing the number of data points by a factor of nine. The resulting averaged images for all H$\alpha$ passbands are shown in Fig. \ref{fig:fig1}(b-l).

Doppler velocities were derived using center-of-mass (COM) centroid method applied to the H$\alpha$ spectral profile \citep{2003ApJ...592.1225U}. The centroid of the wavelength offset computed from a line-depression–type weighting of the profile,

\begin{equation}
\Delta\lambda_{\rm c}
=
\frac{\int (\lambda-\lambda_0)\, w(\lambda)\, \mathrm{d}\lambda}
     {\int w(\lambda)\, \mathrm{d}\lambda}
\end{equation}

where $\lambda_0$ is the rest wavelength of H$\alpha$ and $w(\lambda)$ is a normalized line-depression weighting constructed from the observed profile. The corresponding Doppler velocity is then given by

\begin{equation}
v_{\mathrm{LOS}}
=
\frac{c}{\lambda_0}\,\Delta\lambda_{\rm c} ,
\end{equation}

In practice, the wavelength centroid is computed from the first moment of the normalized line-depression profile with respect to the rest wavelength. As a result, the measured Doppler signal is most sensitive to chromospheric motions and represents a weighted average over multiple wavelengths and formation heights, rather than a velocity at a single, well-defined layer. The GST/VIS H$\alpha$ observations span -1.0 to +1.0 \AA\ from line centre. The far wings ($|\Delta\lambda| \gtrsim 0.6$ \AA) predominantly sample the upper photosphere and temperature-minimum region, the inner wings ($|\Delta\lambda| \approx 0.2$–0.4 \AA) probe the low chromosphere, and the line core originates in the mid-to-upper chromosphere \citep{2012ApJ...749..136L}. Because the line-depression weighting suppresses contributions from the wings, the resulting Doppler velocity is dominated by chromospheric layers, with only a minor influence from lower atmospheric heights.

\begin{figure*}
\centering
\includegraphics[width=.95\linewidth]{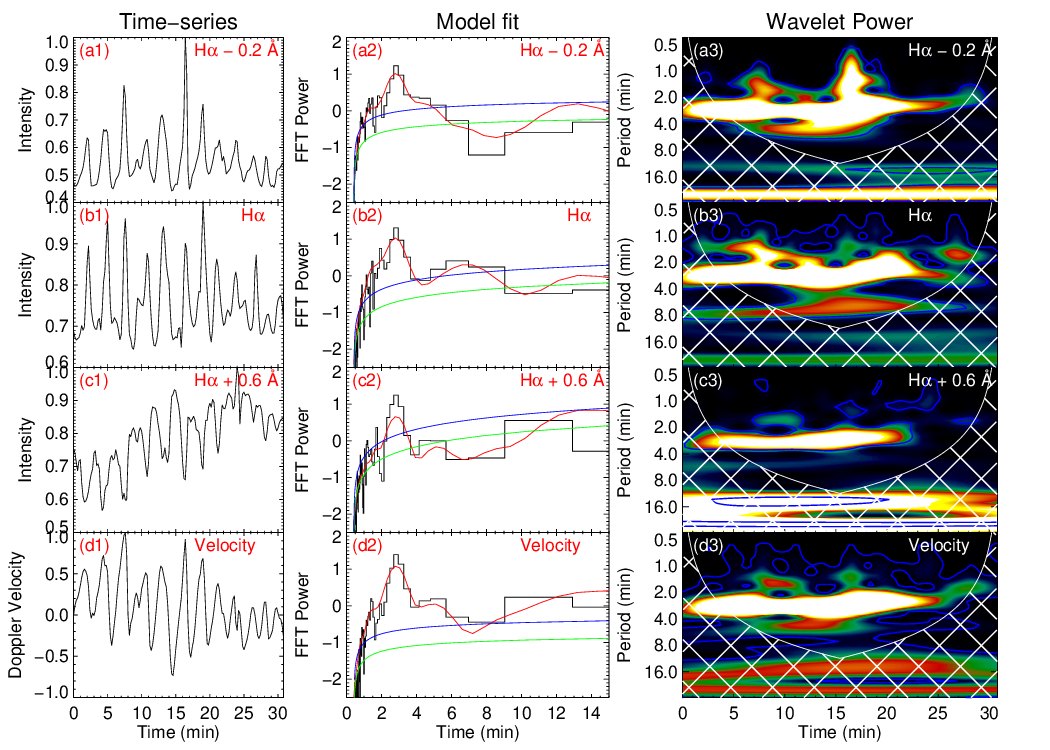}
\caption{Temporal evolution and spectral analysis of H$\alpha$ signals. 
(a1) Intensity of H$\alpha$-0.2\,\AA; 
(b1) Intensity of the H$\alpha$ line core; 
(c1) Intensity of H$\alpha$+0.6\,\AA; 
(d1) Doppler velocity of the H$\alpha$ line. 
All signals are extracted from a pixel at (x = 119$^{\prime\prime}$, y = 159$^{\prime\prime}$); 
(a2--d2) FFT power (black), global wavelet power (red), power-law fit (green), and 95\% local significance level (blue) for the respective time-series; 
(a3--d3) Wavelet transforms of the respective signals.}

\label{fig:fig2}
\end{figure*}

For the analysis of umbral oscillations, we selected all pixels within the white box (about 3000 in total). From these, intensity time-series were extracted for the H$\alpha$ line core and its wings (H$\alpha\pm$0.2 \AA, H$\alpha\pm$0.4 \AA, H$\alpha\pm$0.6 \AA, H$\alpha\pm$0.8 \AA, and H$\alpha\pm$1 \AA). Representative examples of intensity time-series for H$\alpha$-0.2 \AA, the line core, and H$\alpha$+0.6 \AA\ at a single pixel location are shown in Fig. \ref{fig:fig2}(a1–c1). The Doppler velocity time-series at the same pixel is displayed in Fig. \ref{fig:fig2}(d1).

We applied wavelet analysis to these time-series to determine their periodicities, as detailed in sub-section \ref{sec:wavelet}. In addition, the phase relationship between intensity and velocity signals was investigated using cross-wavelet analysis, described in sub-section \ref{sec:cross}.

\section{Methodology}
\label{sec:methodology}

\subsection{Wavelet analysis}
\label{sec:wavelet}

We applied wavelet analysis to study periodicities in the intensity and velocity time-series, which allows us to examine oscillatory power simultaneously in both time and frequency domains. As described by \citet{1998BAMS...79...61T}, this method involves convolving the time-series with a chosen ``mother'' function. Among the commonly used options, e.g., the derivative of a Gaussian (DOG), Paul, and Morlet. We selected the Morlet function, which is essentially a plane wave modulated by a Gaussian. The wavelet transform produces a two-dimensional complex array and the wavelet power is given by the square of its absolute value. The right panel of Fig. \ref{fig:fig2} shows the results of wavelet power spectrum of different time-series. The cross-hatched region marks the ``Cone-of-Influence'' (COI), where edge effects dominate and power values are unreliable. To avoid false detections, we excluded COI values from our analysis. As pointed out by \citet{2016ApJ...825..110A}, detrending can sometimes introduce spurious periodicities, so we did not apply detrending before wavelet transformation.
To assess the significance of oscillatory power, we required a 95\% local confidence level, which depends on accurate background noise modeling. While red and white noise models are often used \citep[e.g.,][]{1998BAMS...79...61T}, we instead adopted a more flexible power-law model with a constant component \citep[e.g.,][]{2016ApJ...825..110A}. This model was fitted to the FFT of each time-series and then used the fitted power law model as a background model. Full details of this method are described in our earlier work \citet{2022MNRAS.517..458S,2024ApJ...966..187S}. The middle panel of Fig. \ref{fig:fig2} illustrates the fitted background model (green), the 95\% confidence level (blue), and the global wavelet power (red). Here, in the global wavelet power, the peaks that exceed the 95\% local confidence level are an indication of significant power and hence significant period. In the right panel of Fig. \ref{fig:fig2}, we overlay 95\% local confidence levels on the wavelet power spectra, and the power lying within these confidence levels is essentially considered significant. In both the wavelet spectra (a3-d3) and global power spectra (a2–d2), we detect significant power near the 3-minute period band. These 3 minute oscillations are well recognized as manifestations of slow magnetoacoustic waves \citep[e.g.,][]{2006RSPTA.364..313B}. These waves travel at the local sound speed and have a periodicity of several minutes, guided via magnetic field lines \citep[e.g.,][]{2006RSPTA.364..447R}. In a sunspot's umbra, when plasma $\beta$ is low, slow magnetoacoustic waves manifest themselves as compressive perturbations essentially oriented along the magnetic field lines. These waves may be excited as upward-propagating waves or may form standing waves due to reflections at boundaries created by steep temperature gradients between the upper photosphere and the transition region \citep[e.g.,][]{2008SoPh..251..501Z}. We further investigate these chromospheric oscillations by performing a phase-difference analysis between the intensity time-series of the H$\alpha$ passbands (core, blue wing, red wing) and the velocity time-series of the H$\alpha$ line.

\begin{figure*}
  	\mbox{
   \centering
   	\includegraphics[width=.95\linewidth]{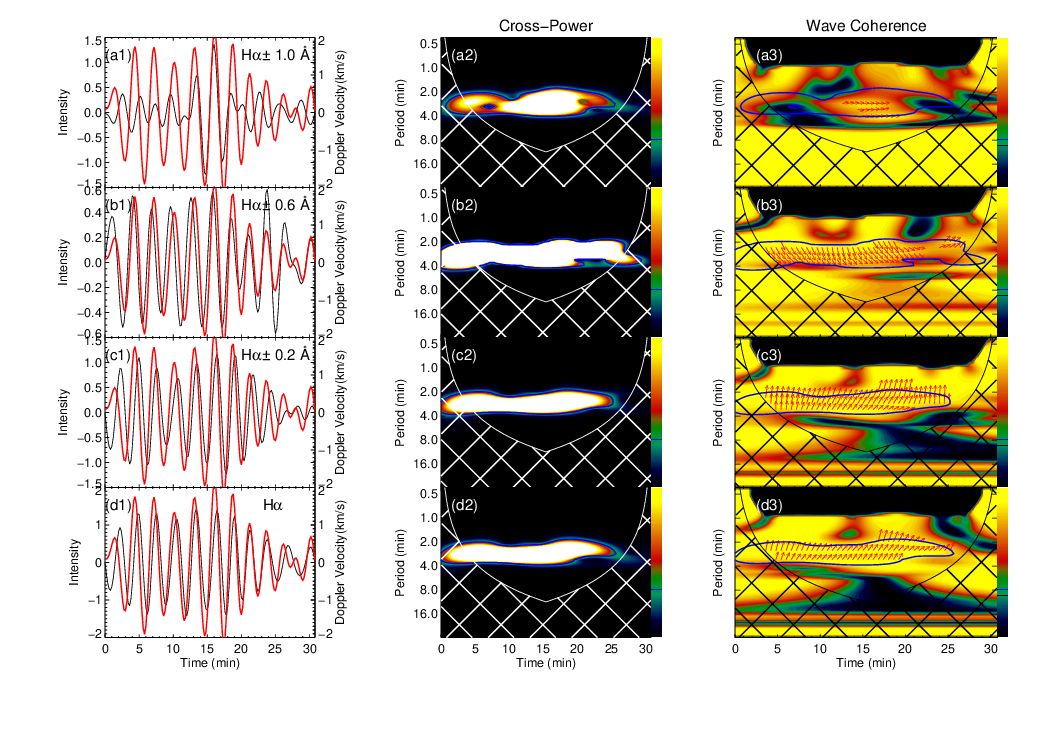} 
   	}
  \caption{Cross-wavelet analysis between intensity of different H$\alpha$ passbands and Doppler velocity (center-of-mass) of H$\alpha$ line. 
(a1) Wing integrating intensity time-series for H$\alpha\pm$ 1.0 \AA\ (black) and Doppler velocity (red);
(b1) Wing integrating intensity time-series for H$\alpha\pm$ 0.6 \AA\ (black) and Doppler velocity (red);
(c1) Wing integrating intensity time-series for H$\alpha\pm$ 0.2 \AA\ (black) and Doppler velocity (red);
(d1) Intensity time-series for H$\alpha$ line core (black) and Doppler velocity (red);
(a2-d2) Cross-wavelet power between the respective H$\alpha$ intensity and Doppler velocity oscillations, with blue contours showing the 95\% local significance level (white noise model);
(a3-d3) Wavelet coherence between intensity and Doppler velocity, with blue contours showing the 95\% local significance level. Red arrows represent relative phase difference: right (0\degree) and left (180\degree) denote propagating waves, vertical ($\pm$ 90\degree) indicate standing waves.}
  \label{fig:fig3}
\end{figure*}

\subsection{Cross-wavelet analysis}
\label{sec:cross}

To investigate correlations and phase differences between intensity and velocity oscillations, we employed cross-wavelet analysis \citep[e.g.,][]{1998BAMS...79...61T}. The cross-wavelet spectrum identifies regions where the two time-series share power, while wavelet coherence reveals areas where oscillations are synchronised, independent of shared power. Coherence values range from zero (no correlation) to one (perfect synchronisation). As shown in Section \ref{sec:wavelet}, the intensity time-series of the H$\alpha$ line core and wings, along with the Doppler velocity of H$\alpha$, exhibit a dominant 3-minute periodicity, consistent with previous observations of chromospheric umbrae \citep[e.g.,][]{2000A&A...354..305C, 2006SoPh..238..231K, 2022ApJ...924..100C}.

To explore the origin of the 3-minute umbral oscillations, we first pre-processed the intensity and velocity time-series. A narrowband filter spanning 2–4 minutes was applied to all H$\alpha$ passbands as well as to the Doppler velocity time-series. To investigate phase relationships, we performed a cross-wavelet analysis between the intensity time-series of different H$\alpha$ passbands (line core and wings) and the centroid-derived H$\alpha$ Doppler velocity time-series at each pixel within the selected region. Because the Doppler velocity is core-weighted and height-integrated, it can be treated as a common chromospheric reference signal, representing upper-chromosphere-weighted dynamics rather than a discrete formation height.

For the wing measurements, we first computed the integrated intensity, $\Delta$ I(t) = (I$_{\mathrm{r}}$(t) + I$_{\mathrm{b}}$(t))/2, where I$_{\mathrm{r}}$ and I$_{\mathrm{b}}$ denote the intensity in the red and blue wings of the spectral line, respectively sampled at equal wavelength offsets from the line center. This symmetric combination was used to suppress Doppler contributions, since the velocity response of chromospheric line wings is antisymmetric. In this way, we isolate thermodynamic signatures while minimizing Doppler-induced antisymmetry \citep{1983A&A...123..263V,2006ASPC..354..313U, 2011ApJ...736...71W}. Examples of the resulting filtered integrated-intensity signals are shown in Fig. \ref{fig:fig3} (panels a1–c1) in black, while the filtered Doppler velocity is shown in red. These examples correspond to a single spatial location; however, the same preprocessing was applied to all intensity and Doppler velocity time series prior to the subsequent analysis.

Cross-wavelet power is computed by multiplying the wavelet transform of the integrated intensity signal with the complex conjugate of the velocity transform and then taking the squared magnitude. The results for representative wing combination (H$\alpha$ $\pm$ 1.0 \AA, H$\alpha$ $\pm$ 0.6 \AA, H$\alpha$ $\pm$ 0.2 \AA, and line core) at a sample location (x=119$^{\prime\prime}$, y=159$^{\prime\prime}$) are shown in Fig. \ref{fig:fig3} (a2–d2). To quantify synchronisation, the cross-wavelet power was normalised by the product of the individual power spectra, producing the wavelet coherence maps displayed in Fig. \ref{fig:fig3} (a3–d3). The local significance levels 95\% were calculated using a white noise model \citep[e.g.,][]{1998BAMS...79...61T}, are overplotted in blue, with significant power defined as within these contours. 

Phase differences between the intensity and velocity signals were derived from the real and imaginary components of the cross-wavelet power. To ensure reliability, we applied strict selection criteria: (1) the cross-power must lie within the 95\% local significance region, (2) values inside the Cone of Influence (COI) were excluded to avoid edge effects, and (3) only regions with coherence above 0.7 were considered. Applying these criteria to all 3000 umbral pixels allowed us to extract consistent phase differences between the wing-integrated intensity and the centroid-derived H$\alpha$ Doppler velocity time-series, as well as between the H$\alpha$ line-core intensity and the centroid-derived H$\alpha$ Doppler velocity time-series. For a representative case, the phase differences between intensity and Doppler velocity are shown in the form of phase arrows overlaid on the wavelet coherence power map in Fig. \ref{fig:fig3} (panels a3–d3). We define the phase difference as $\phi = \phi_I - \phi_V$, such that a positive phase indicates that intensity lags velocity. Positive Doppler velocities indicate motion toward the observer (blueshift), while negative velocities indicate motion away from the observer (redshift). Arrows pointing to the right indicate a phase difference of 0\degree, arrows pointing to the left indicate 180\degree, and vertical arrows (upward or downward) indicate a 90\degree phase difference. These phase differences were used to analyse the mode of the umbral oscillations. Phase analysis is discussed in detail in Section \ref{sec:results}. Phase differences were estimated for all pixels, and the statistical behaviour of the oscillations within the selected umbral region is also presented in Section \ref{sec:results}.

\section{Results}
\label{sec:results}

\begin{figure*}
  	\mbox{
   \centering
   	\includegraphics[width=.95\linewidth]{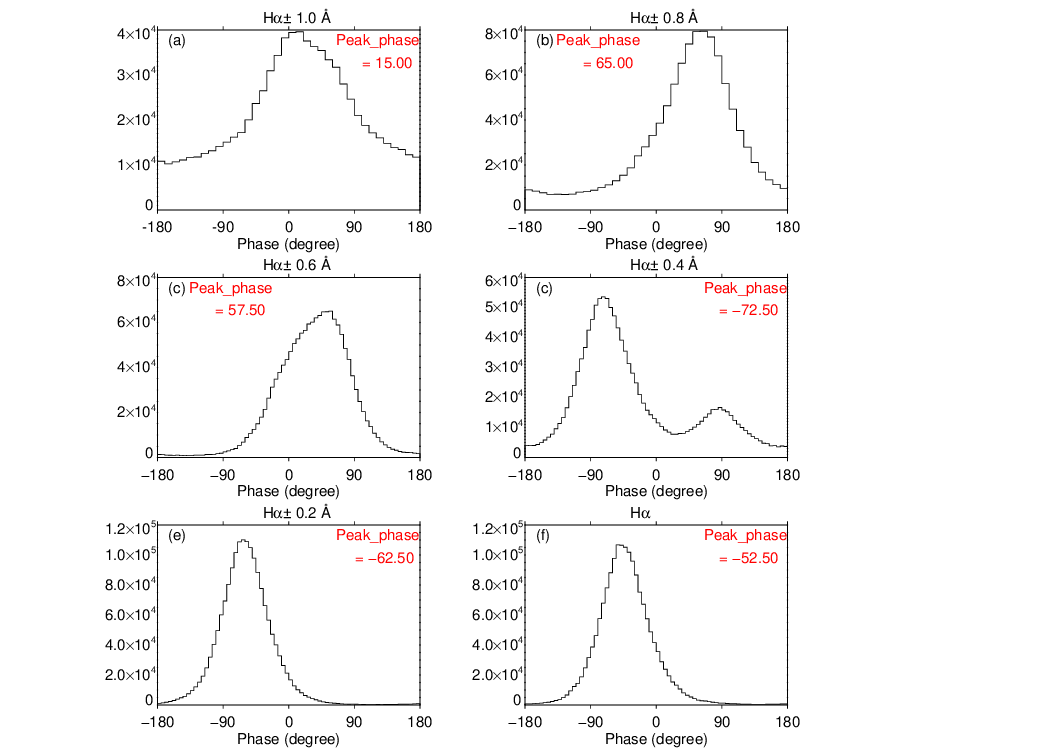} 
    }
  \caption{(a–e) Histograms of significant phase differences between H$\alpha$ intensity (averaged from both line wings) and Doppler velocity (derived from line center-of-mass) time-series at selected umbral locations.
(f) Histogram of significant phase differences between H$\alpha$ line-core intensity and Doppler velocity (derived from line center-of-mass) time-series at the same locations. The peak phase difference are menioned in each panel.}
  \label{fig:fig4}
\end{figure*}

According to linear MHD theory, slow magnetoacoustic waves are compressive in  nature and appear as periodic disturbances in both intensity and Doppler velocity, propagating at speeds close to the local sound speed. The phase difference between velocity and density (and therefore intensity) perturbations depends on the wave type: in-phase or out-of-phase (0\degree\ or 180\degree) are expected for propagating waves under linear, adiabatic conditions, while a quarter-period phase shift ($\pm90\degree$) corresponds to standing waves \citep[e.g.,][]{1990A&A...236..509D,2009ApJ...696.1448W,2013A&A...551A.137M}. Measuring the phase relationship between intensity and Doppler velocity oscillations is thus a key diagnostic for distinguishing wave modes in the solar atmosphere, particularly in the chromosphere and corona. Several studies have employed this approach to differentiate between propagating and standing slow waves \citep[e.g.,][]{2002SoPh..209..265S,2003A&A...406.1105W,2009ApJ...696.1448W,2010ApJ...721..744K,2010ApJ...713..573M}. For instance, \citet{2009ApJ...696.1448W} analyzed Hinode/EIS observations and found intensity oscillations leading Doppler velocity by 20\degree-30\degree, consistent with upward-propagating waves. Similarly, \citet{2003A&A...402L..17W} reported standing waves in hot coronal loops observed with SOHO/SUMER, with a phase shift of 90\degree. These observational phase differences align with linear MHD predictions: upward-propagating waves show velocity and density nearly in phase, while standing waves exhibit a quarter-period lead of density over velocity. In this study, we compare the theoretically expected phase differences with the observed phase differences between intensity and Doppler velocity oscillations to interpret our observational results. 

In Fig. \ref{fig:fig3}, the phase difference shows an approximately in-phase relationship between the intensity and velocity signals at the H$\alpha$ $\pm$ 1.0 \AA\ wing. This is often interpreted as consistent with upward-propagating waves. At higher atmospheric layers, the direction of the phase arrows changes, showing phase offsets closer to $\pm 90^\circ$, which are often interpreted as indicative of standing wave. Phase differences were also computed at each spatial location from the real and imaginary components of the cross-wavelet power \citep{1998BAMS...79...61T}, following the selection criteria described in Section~\ref{sec:cross}. Histograms of the phase differences for the different cases are shown in Fig.~\ref{fig:fig4}. Panels (a)–(e) correspond to the wing, while panel (f) shows the line-core results. In panel (a), the phase differences peak near $15^\circ$, whereas in the remaining panels, which correspond to chromospheric passbands, the distributions peak closer to $\pm 90^\circ$. These phase differences are suggestive of propagating waves in the layers sampled by the H$\alpha$ $\pm$ 1.0~\AA\ far wings, and standing wave at chromospheric heights sampled by the H$\alpha$ $\pm$ 0.4~\AA, H$\alpha$ $\pm$ 0.2~\AA, and the H$\alpha$ line core \citep[e.g.,][]{2009ApJ...696.1448W,2010ApJ...713..573M}. Correspondingly, within the 3-minute band, the phase arrows increasingly exhibit a quarter cycle phase shift ($\pm 90^\circ$) between intensity and Doppler velocity at chromospheric heights, suggesting a standing wave in the umbral chromosphere. However, because the intensity and velocity diagnostics may not sample the same atmospheric height, these I--V phase differences are therefore not used alone as definitive evidence for standing waves.

\begin{figure*}
  	\mbox{
   \centering
   	\includegraphics[width=.95\linewidth]{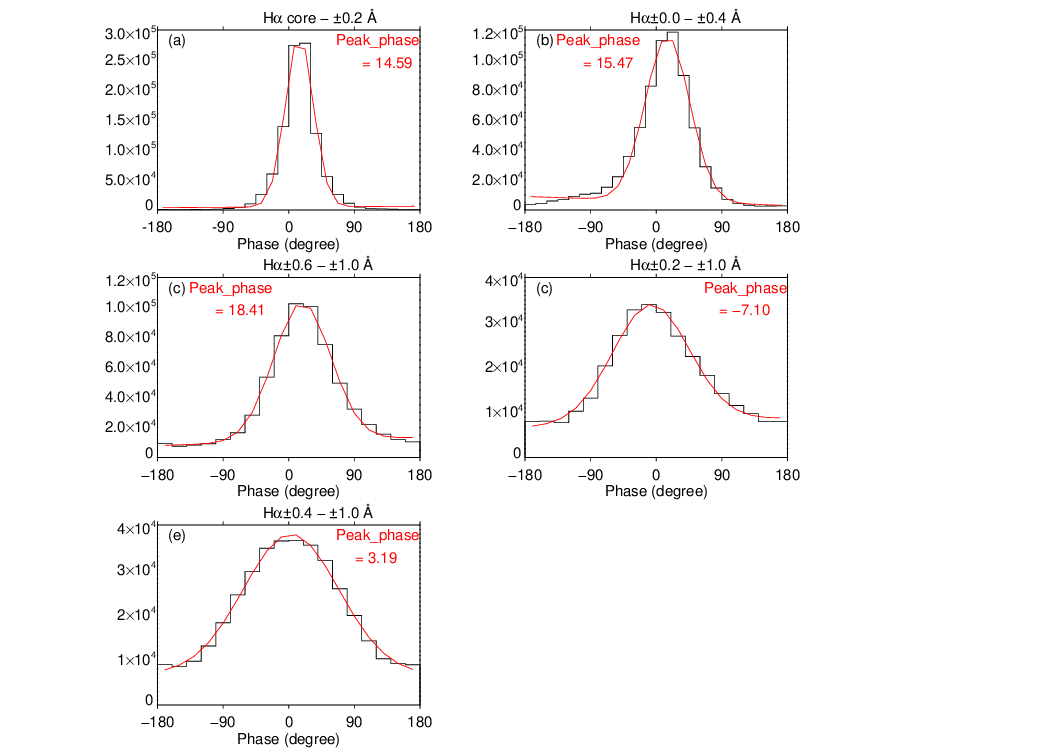} 
    }
  \caption{Histogram of phase differences between intensity signals from different combinations of H$\alpha$ passbands in the 3 min band. The specific passband pairs are indicated at the top of each panel. The red curves represent Gaussian fits to the distributions, and the corresponding peak phase values are indicated in each panel. The phase differences are predominantly clustered around 0\degree, with small deviations (within $\pm$20\degree), indicating strong phase coherence between the intensity oscillations sampled by the different H$\alpha$ passbands.}
  \label{fig:fig5}
\end{figure*}

To address this limitation, we additionally analyse the phase differences between intensity signals measured at different H$\alpha$ passbands. We consider the following passband combinations: H$\alpha$ line core -- $\pm$0.2 \AA, H$\alpha$ line core -- $\pm$0.4 \AA, H$\alpha$ $\pm$0.2 -- $\pm$1.0 \AA, H$\alpha$ $\pm$0.4 -- $\pm$1.0 \AA, and H$\alpha$ $\pm$0.6 -- $\pm$1.0 \AA. These passbands sample different parts of the H$\alpha$ profile and are therefore expected to have different atmospheric contributions. For each pair, we compute the cross power between the corresponding intensity time series and derive the phase difference from the real and imaginary components of the cross spectrum. The analysis is performed for all pixels within the selected umbral region. The resulting phase-difference distributions are shown in Fig.~\ref{fig:fig5}. Each distribution is fitted with a one-dimensional Gaussian function, and the fitted peak phase is indicated in the corresponding panel. The phase distributions are predominantly concentrated close to 0\degree, with fitted peak values lying within approximately $\pm$20\degree. This indicates that the H$\alpha$ intensity oscillations sampled by these passband combinations are largely phase coherent in the 3-minute band. We do not find a systematic increase or decrease of phase difference with increasing passband separation. The absence of a clear monotonic phase variation with increasing passband separation suggests that the observed oscillations are not consistent with a simple picture of purely upward-propagating waves between well-separated atmospheric layers \citep[e.g.,][]{2006ApJ...640.1153C, 2017ApJ...847....5K}. Instead, the near-zero phase differences imply that the sampled chromospheric layers oscillate nearly in phase over the range of atmospheric contributions probed by the H$\alpha$ line. This predominantly in-phase relation is consistent with slow magnetoacoustic oscillations exhibiting a significant standing or partially standing component in the umbral chromosphere. The small deviations from 0\degree may reflect radiative transfer effects or indicate the presence of a weak propagating component.

In addition to their use in characterizing wave mode, phase shifts can also provide insight into damping processes that influence the propagation and evolution of slow waves. Previous studies have shown that radiative cooling, thermal conduction, compressive viscosity, and imbalances between heating and cooling can introduce measurable phase shifts between velocity and density oscillations \citep[e.g.,][]{2009A&A...494..339O, 2021SoPh..296...20P, 2022SoPh..297....5P}. The phase differences identified in this study, when considered together with theoretical expectations and earlier observational results, suggest that multiple wave modes may coexist in the chromosphere. Their interpretation should account not only for ideal wave properties but also for dissipative processes that can modify phase relations and amplitudes during propagation. In Section \ref{sec:discussion}, we further examine these phase differences between the different H$\alpha$ passbands and discuss their compatibility with a chromospheric resonant cavity model.

\section{Discussion and Conclusion} \label{sec:discussion}

We analyzed high-resolution observations from the Goode Solar Telescope at Big Bear Solar Observatory (GST/VIS) to investigate oscillations in the umbral region of a strongly magnetized sunspot. Our study focused on intensity oscillations in both the core and wings of the H$\alpha$ line, as well as Doppler velocity oscillations derived from the same line. Statistical analysis revealed a clear dominance of three-minute oscillations at multiple locations within the chromospheric umbra. Furthermore, phase spectra analysis reveals phase differences compatible with upward propagating waves in the lower chromosphere and standing or partial standing waves at higher chromospheric layer. These findings are consistent with the chromospheric resonator model, in which three-minute oscillations are interpreted as slow magnetoacoustic waves resonating within a cavity bounded by the photosphere and the transition region \citep[e.g.,][]{1981SvAL....7...25Z,2008SoPh..251..501Z}. Steep temperature gradients at the cavity boundaries act as partial reflectors, with the transition region reflecting upward-propagating waves downward and the temperature minimum reflecting waves back upward, thereby establishing a resonant chromospheric cavity that sustains the observed three-minute oscillations \citep[e.g.,][]{2007AstL...33..622Z,2008SoPh..251..501Z,2011ApJ...728...84B,2015A&A...580A.107S,2016ApJS..224...30Y,2019A&A...627A.169F}.

In this study, we present new observational results that provide additional support for the chromospheric resonator model in sunspots. We utilized the high spatial and spectral resolution of BBSO/GST VIS data. Unlike previous studies limited to a narrow range of spectral lines \citep[e.g.,][]{2020NatAs...4..220J}, the VIS spectrograph allows simultaneous observations across multiple H$\alpha$ passbands, enabling us to probe a wide range of chromospheric heights. This multi-height coverage provides a more detailed view of wave propagation and resonant phenomenon in sunspot chromospheres. Using a single centroid-derived Doppler velocity together with intensities from multiple H$\alpha$ passbands allows us to examine phase differences across atmospheric height in a consistent way. This approach makes it possible to compare velocity–intensity phase relationships from the photosphere to the chromosphere within one analysis. The phase progression from the far wings to the line core shows a systematic change across H$\alpha$ passbands. At lower heights, the phase differences are compatible with predominantly upward wave propagation. At chromospheric heights, the phase offsets shift toward values commonly associated with standing wave in umbral 3-minute oscillations. Such a pattern is qualitatively consistent with waves propagating upward from lower layers, undergoing partial reflection, and forming standing or partially standing wave patterns higher in the atmosphere. Unlike earlier studies that examined phase differences as a continuous function of frequency \citep{1986ApJ...301..992L}, we focus on the statistical distribution of cross-wavelet phase within selected frequency bands. This approach is better suited to the intermittent and non-stationary nature of chromospheric oscillations and allows dominant phase relationships within the 3-minute band to be identified. The interpretation of chromospheric phase relations requires caution, as Doppler velocities in chromospheric lines are influenced by both velocity and thermodynamic effects, particularly under shock conditions \citep{2013SoPh..288...73M}.

The phase change from H$\alpha$ $\pm$ 0.6 \AA\ to H$\alpha$ $\pm$ 0.4 \AA\ indicates a rapid transition in the phase difference between the corresponding atmospheric layers. Such phase changes can occur when the sampled layers approach velocity or temperature nodal regions in a chromospheric resonant cavity, where phase discontinuities of 180\degree\ are expected \citep{2020ApJ...900L..29F}. However, the absence of a clear 180\degree\ phase jump and the broad, partially overlapping formation ranges of the H$\alpha$ passbands prevent a definitive identification of a nodal layer. The combined intensity–velocity and intensity–intensity phase analyses support a scenario that cannot be explained solely by simple upward propagation. Instead, the observations are compatible with a mixed wave mode involving partially standing slow magnetoacoustic oscillations within the chromosphere. Such behaviour is qualitatively consistent with expectations from chromospheric resonator models. 

At higher chromospheric layers, the H$\alpha$ line core exhibits a phase difference of -52.50\degree, likely reflecting a standing or partially standing component or a superposition of upward and downward propagating waves. This observation is comparable to the phase shift of $\pi$/4 reported by \citet{2016ApJS..224...30Y}, and is broadly consistent with expectations for resonant cavity models. \citet{1986ApJ...301.1005L} measured intensity from chromospheric line-core diagnostics (Ca II H emission and He I absorption) and demonstrated systematic phase shifts between intensity and Doppler velocity as a function of frequency, with characteristic phase lags of 70\degree\ in the three-minute band. While our Halpha observations use fixed passband intensities rather than spectroscopic line-core measurements, shows qualitatively similar trends, consistent with chromospheric resonant model. Additionally, we do not observe any sawtooth profiles in intensity or Doppler velocity time-series, indicating that the oscillations are predominantly harmonic rather than shock-dominated. Overall, our results reproduce several key features predicted by numerical simulations, including the simultaneous presence of propagating and standing waves, and provide observational support consistent with the interpretation that the sunspot chromospheric cavity resonates at three-minute periods. Moreover, the findings suggest the possibility of upward leakage of slow waves into the corona, which may contribute to energy transfer within the solar atmosphere.

Previous studies have shown that the chromosphere can act as a leaky resonator, partially reflecting slow magnetoacoustic waves at the transition region while allowing higher-frequency waves to propagate into the corona. The characteristics of coronal line-of-sight frequency spectra are influenced by the size of the chromospheric cavity: larger cavities produce narrow and dynamic spectra, whereas smaller cavities generate noisier spectra with pronounced peaks \citep[e.g.,][]{2015A&A...580A.107S, 1987ApJ...312..457T, 2012ApJ...746..119R}. Numerical simulations further indicate that broadband velocity perturbations in the chromosphere can generate standing waves within the chromosphere and propagating wave trains in the corona, leading to dominant three-minute oscillations without requiring a narrowband driver \citep[e.g.,][]{2011ApJ...728...84B, 2002ESASP.505..183H, 2009A&A...494..339O}. These mechanisms serve as a theoretical framework for understanding the origin of upward and downward propagating waves within the sunspot umbra. We observe upward propagating waves and standing or partially standing waves, across multiple chromospheric heights. These waves provide measurable properties such as phase differences, periods, and frequency content that can serve as important input for sunspot seismology. Previous numerical studies \citep[e.g.,][]{2008SoPh..251..501Z, 2008SoPh..251..523T, 2015A&A...580A.107S} show that parameters including cavity size, boundary conditions, and cut-off frequencies strongly influence wave propagation, reflection, and resonance. Although direct comparisons with models were not conducted here, combining our observations with simulations can help constrain cavity dimensions, vertical stratification, and atmospheric structure. This approach demonstrates how sunspot seismology can be used to probe the internal structure and evolution of sunspot atmospheres, highlighting the importance of combining observational data with theoretical models.

Our high-resolution BBSO/GST observations are broadly consistent with the chromospheric resonator model, revealing the coexistence of upward propagating waves, as well as standing or partially standing waves, across multiple chromospheric heights within sunspot umbrae. Future high-resolution facilities, such as the Daniel K. Inouye Solar Telescope (DKIST) \citep{2020SoPh..295..172R} and the European Solar Telescope (EST) \citep{2022A&A...666A..21Q}, with their multi-line spectroscopy and sensitive Doppler capabilities, will allow more detailed studies of these phenomena. Such observations will help constrain cavity properties, track wave propagation and energy transfer, and strengthen sunspot seismology by linking precise observations with numerical models.
\section*{Acknowledgements}

K.S., L.F., and D.Y. is supported by National Natural Science Foundation of China (NSFC, 11803005), the Guangdong Natural Science Funds for Distinguished Young Scholar (2023B1515020049), the Shenzhen Science and Technology Project (JCYJ20240813104805008), and the Science and Technology Program of Guangdong Province (grant No. 2025B1212050001) and the Specialized Research Fund for State Key Laboratory of Solar Activity and Space Weather. The authors acknowledge Prof. Wenda Cao for providing the GST observational data used in this study.

\medskip
\noindent \textbf{Use of AI Tools:} The authors used AI-assisted language tools to improve the clarity and readability of the manuscript. They carefully reviewed and edited all content themselves and take full responsibility for the final version.

\section*{Data Availability}
The authors appreciate the use of data from the Goode Solar Telescope (GST) of the Big Bear Solar Observatory (BBSO). The operation of BBSO is supported by the US NSF AGS-2309939 grant and NJIT. GST operation is partly supported by the Korea Astronomy and Space Science Institute and the Seoul National University. 
 







\bsp	
\label{lastpage}
\end{document}